\documentclass{article}
\usepackage{spconf2,amsmath,graphicx}
\usepackage{amsfonts}
\usepackage{bm}
\usepackage{hyperref}
\usepackage{booktabs}
\usepackage{tabularx}

\usepackage{multirow}

\title{A Style Transfer Approach to Source Separation}

\name{Shrikant Venkataramani$^{\sharp}$ \qquad 
      Efthymios Tzinis$^{\flat}$ \qquad 
      Paris Smaragdis$^{\flat, \natural,}$\sthanks{Supported by NSF grant \#1453104}}
\address{$^{\sharp}$University of Illinois at Urbana-Champaign, Department of Electrical and Computer Engineering \\
         $^{\flat}$University of Illinois at Urbana-Champaign, Department of Computer Science \\
         $^{\natural}$Adobe Research}

\begin{document}
\ninept
\maketitle
\begin{abstract}
Training neural networks for source separation involves presenting a mixture recording at the input of the network and updating network parameters in order to produce an output that resembles the clean source. Consequently, supervised source separation depends on the availability of paired mixture-clean training examples. In this paper, we interpret source separation as a style transfer problem. We present a variational auto-encoder network that exploits the commonality across the domain of mixtures and the domain of clean sounds and learns a shared latent representation across the two domains. Using these cycle-consistent variational auto-encoders, we learn a mapping from the mixture domain to the domain of clean sounds and perform source separation without explicitly supervising with paired training examples. 
\end{abstract} 
\section{Introduction}
\label{sec:intro}
With the recent advances in deep learning, several sophisticated neural networks have been proposed and used for source separation~\cite{luo2018tasnet, shi2019furcanet, hershey2016deep, venkataramani2018end, luo2017deep, wisdom2019differentiable}. Most of these efforts have focused on solving the problem in a supervised setting where there is access to the clean version of the sources in the mixture. The mixture is given as an input to the network and the network is trained to produce an output that resembles the clean sounds. Consequently, supervised source separation using neural networks relies on the availability of paired mixture-clean data in the training set and cannot be used when such paired datasets are not available or expensive to collect. To relax these constraints, a few recent papers use other forms of information like the spatial separation between the sources in a multi-microphone setting, to train the networks for unsupervised source separation~\cite{tzinis2019unsupervised, seetharaman2019bootstrapping, drude2019unsupervised, drude2019unsupervisedbeamforming}. However, these constraints continue to impose restrictions on single-channel source separation, where such secondary forms of information about the sources are not available. 

The goal of style transfer (aka. domain translation) is to map a data-point from the source domain to a corresponding data-point in the target domain. During this mapping, the content remains unchanged while the style of the data undergoes a modification. In the case of audio, several problems of interest fall under the category of domain translation. For example, in voice conversion~\cite{kameoka2018stargan} a sentence recorded by a speaker is transformed to make it sound as if the sentence was spoken by someone else. In source separation, the network learns to map the recording of a mixture to the clean recording of the source we wish to extract~\cite{luo2018tasnet, shi2019furcanet, hershey2016deep, venkataramani2018end, luo2017deep, wisdom2019differentiable}. The mappings from the source to the target domain can be learned in a supervised or an unsupervised manner. Unsupervised Domain Translation~(UDT) attempts to learn the mappings from the source to the target domain, without paired training examples. Accordingly, using UDT for source separation would relax the constraints on the training data and enable us to extract the desired source from the mixture without paired training examples. 

In this paper, we relax the constraints on the training data by interpreting single-channel source separation as an unsupervised style-transfer problem. Instead of paired mixture-clean training examples,  we assume that we have several mixture recordings and several clean examples of the source we wish to isolate. However, we do not have any information about the relationships between the clean recordings of the source and the mixture recordings. In fact, the mixture and source examples could be completely unpaired and come from distinct recordings. We investigate the use of UDT where we learn a shared latent space for the mixtures and the clean sounds. The joint latent representation enables us to learn a mapping from the domain of mixtures to the domain of clean examples, performing source separation without supervising with paired training examples. Finally, we emphasize that we do not define style and content explicitly by imposing specific modeling structures. The neural network proposed automatically understands these aspects based on the training examples provided for the two domains and can possibly be extended to other UDT applications in audio.


\section{Related Work}
\label{sec:related_work}
Domain translation maps a point in the source domain to a corresponding point in the target domain. As shown in~\cite{liu2017unsupervised}, such domain translation problems can be probabilistically interpreted as learning a joint distribution for the source and target domains. In the case of images, domain transfer algorithms have been used to map images between different painting styles while keeping the content fixed (for eg., Monet paintings to natural images and vice versa). Several Generative Adversarial Network~(GAN) based architectures have been proposed for supervised image-to-image translation~\cite{isola2017image, kim2017learning, choi2018stargan}. 

More recently, GANs have also been successfully used for unsupervised image-to-image translation~\cite{zhu2017unpaired}. To simplify the complexity of GAN based networks, Liu et.al., proposed the UNIT algorithm~\cite{liu2017unsupervised} using Variational Auto-Encoders~(VAEs) to learn generative models for the source and target domains. As described in \cite{liu2017unsupervised}, the source and target domain training examples define a marginal distribution for the source and target domains respectively. Further, the goal of UDT is to estimate the joint distribution of the two domains, given their marginal distributions. This problem is an ill-posed problem and requires additional assumptions on the joint distribution~\cite{lindvall2002lectures}. To this end, Liu et. al., show that a necessary condition for UDT is to have a shared latent space between the source and target domains. This implies that a pair of points having the same content in the source and target domains should be mapped to the same point in the latent space. Since we do not have access to such paired training examples, we use additional consistency terms in the cost-function to enforce a shared latent space as described in Section~\ref{ssec:cost_function}.

In the case of style transfer problems for audio data, style and content are ambiguous~\cite{grinstein2018audio}. Consequently, style transfer approaches for audio have largely resulted in efforts that match the timbres of the source and target domains (for example, opera sounds have been matched to sound like cats licking milk)~\cite{grinstein2018audio, ramani2018autoencoder}. In such cases, it is difficult to define what a good network output is supposed to sound like. Also, these efforts have not been targeted at any specific application and a tangible metric to evaluate the performance of the network is also difficult to obtain. In the case of music, style transfer approaches have been used for timbral transfer~\cite{huang2018timbretron, huangdeep}. Mor et. al., use audio style transfer to translate music across different instruments and styles and demonstrate that their approach transfers stylistic musical elements also~\cite{mor2018universal}. GAN based networks have been proposed for voice transfer applications~\cite{kameoka2018stargan, kaneko2018cyclegan, kaneko2019cyclegan} and singing voice separation~\cite{michelashvili2019singing} without paired data. In this paper, we explore the use of a network inspired by the UNIT architecture~\cite{liu2017unsupervised} for UDT and show how it can be used for source separation without paired training examples. In doing so, we use a very narrow interpretation of style and content for source separation: the content refers to the source we wish to extract from the mixture and style refers to the presence or absence of interfering sources. The performance of style transfer can also be evaluated using the metrics we commonly use to evaluate source separation performance~\cite{vincent2006performance, le2019sdr}.   

\section{Unsupervised Domain Translation for Source Separation}
\label{sec:unsupervised_domain_translation_ss}
We now introduce the idea of UDT for audio and show how it can be used for source separation. We interpret source separation as a style transfer problem of learning a mapping from the mixture domain to the domain of clean sounds. For the sake of simplicity, we assume that the goal is to extract the female speaker from a mixture of one male and one female speaker respectively. However, the approach developed below is general and can be used to extract any desired source from a mixture if example recordings of the source are available. Thus, in our setting, the source domain training set consists of several example mixtures of male and female speakers. The target domain training set consists of several examples of isolated female speech.

\begin{figure}[t]
\centering
  \includegraphics[clip, trim = 0cm 0cm 0cm 0cm, width=\linewidth]{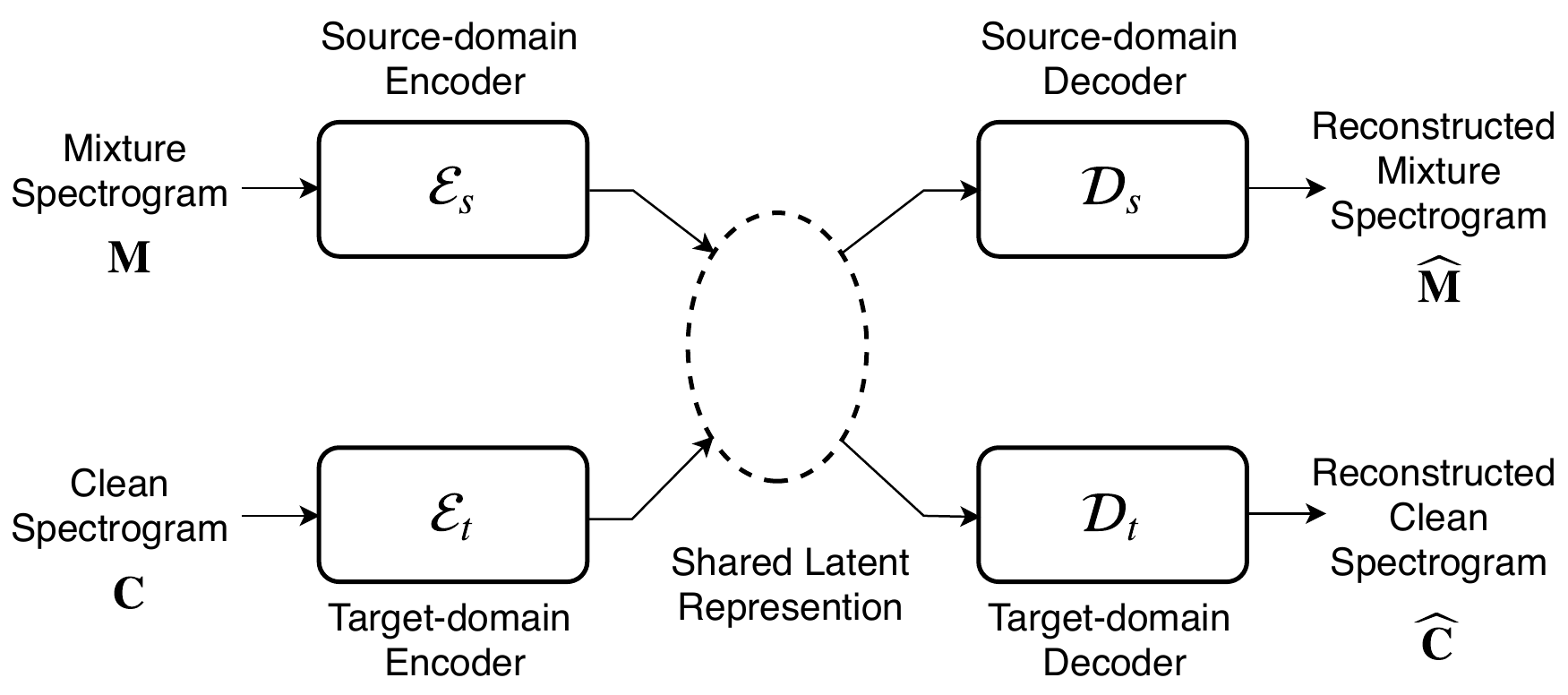}
  \caption{Block diagram of network used for UDT. The network is a combination of two VAEs that share a latent space between them. The clean spectrogram $\mathbf{C}$ comes from an unrelated recording of the source we wish to extract from the mixture spectrogram $\mathbf{M}$. }~\label{fig:udt_block_diagram}
\end{figure}

\subsection{Network Architecture}
\label{ssec:network_architecture}
We first describe the network architecture used for source separation using UDT. We use a modified version of UNIT~\cite{liu2017unsupervised} for our audio domain translation application. Figure~\ref{fig:udt_block_diagram} gives the block diagram of the network architecture used. All the audio signals are first transformed into the time-frequency domain using a short-time Fourier transform (STFT) operation. We compute the magnitude spectrograms from their STFT representations and use them as the inputs to the network. The networks are also trained to estimate magnitude spectrograms at their outputs. From Figure~\ref{fig:udt_block_diagram}, we see that the network basically consists of two individual auto-encoders. The encoder~$\mathbf{\mathcal{E}_{s}}$ and decoder~$\mathbf{\mathcal{D}_{s}}$ act as a VAE for the source domain (domain of mixtures). Similarly, $\mathbf{\mathcal{E}_{t}}$ and $\mathbf{\mathcal{D}_{t}}$ act as an encoder and decoder forming the VAE for the target domain (domain of isolated female speech). The magnitude spectrograms of the mixture and female speech examples are given as inputs to respective VAEs. To implement the VAEs, we use the reparameterization trick~\cite{kingma2013auto}. We add a random vector drawn from the multi-variate Gaussian distribution $\mathcal{N}(0,I)$ to the encoder outputs to backpropagate the gradients through the sampling step.

Instead of allowing the VAEs to operate independently, we force them to learn a shared or joint latent representation for the source (mixtures) and target (female speech) domains. Similar latent space sharing strategies have been previously applied in non-negative matrix partial cofactorization~(NMPCF) for extracting the common source across multiple spectrograms~\cite{kim2011nonnegative}. The implications of a shared latent representation are as follows: consider a mixture-clean pair of spectrograms $(\mathbf{M_{x}}, \mathbf{C_{x}})$ where the mixture is formed by mixing the clean utterance corresponding to the spectrogram $\mathbf{C_{x}}$ with an interfering source. The respective encoders $\mathbf{\mathcal{E}_{s}}$ and $\mathbf{\mathcal{E}_{t}}$ would map $\mathbf{M_{x}}$ and $\mathbf{C_{x}}$ to the same latent representation. Also, given an arbitrary latent representation $h$, we can use the two decoders $\mathbf{\mathcal{D}_{s}}$ and $\mathbf{\mathcal{D}_{t}}$ to generate a pair of corresponding mixture and clean spectrograms. Training these auto-encoders using only reconstruction terms in the cost function results in two independently trained auto-encoders for the source and target domains. To ensure a shared latent space, we train the auto-encoders in tandem and also use cycle-consistency loss terms in the cost-function~\cite{liu2017unsupervised, zhu2017unpaired, kaneko2018cyclegan}. The details of the cost-function used to train the network are given in Section~\ref{ssec:cost_function}. In addition, the final layer of the encoders and the first layer of the decoders also share their weights. 

From the network architecture in Figure~\ref{fig:udt_block_diagram}, we see that we can now simultaneously obtain four types of mapped outputs. Given a mixture spectrogram $\mathbf{M}$, $\mathbf{\mathcal{E}_{s}}( \mathbf{M})$ denotes the latent representation of $\mathbf{M}$. Alternatively, for a source spectrogram $\mathbf{C}$, $\mathbf{\mathcal{E}_{t}}( \mathbf{C})$ returns its latent representation. We now have the following mapping pathways:
\begin{itemize}
  \item The path $\mathbf{\mathcal{E}_{s}} \to \mathbf{\mathcal{D}_{s}}$ gives the mixture reconstruction~$\mathbf{\widehat{M}}$.
  
  \item The path $\mathbf{\mathcal{E}_{s}} \to \mathbf{\mathcal{D}_{t}}$ extracts the source of interest from the mixture.
  
  \item The path $\mathbf{\mathcal{E}_{t}} \to \mathbf{\mathcal{D}_{t}}$ reconstructs the source spectrogram giving ~$\mathbf{\widehat{C}}$.
  
  \item The path $\mathbf{\mathcal{E}_{t}} \to \mathbf{\mathcal{D}_{s}}$ gives a dummy mixture spectrogram that contains $\mathbf{C}$ as one of the sources.
\end{itemize}
At inference time, the mixture spectrogram is transformed to the latent space by $\mathbf{\mathcal{E}_{s}}$. The target decoder $\mathbf{\mathcal{D}_{t}}$ operates on this latent representation and estimates the separated spectrogram of the source.

\begin{figure}[!ht]
\centering
  \includegraphics[clip, trim = 0cm 0cm 0cm 0cm, width=\linewidth]{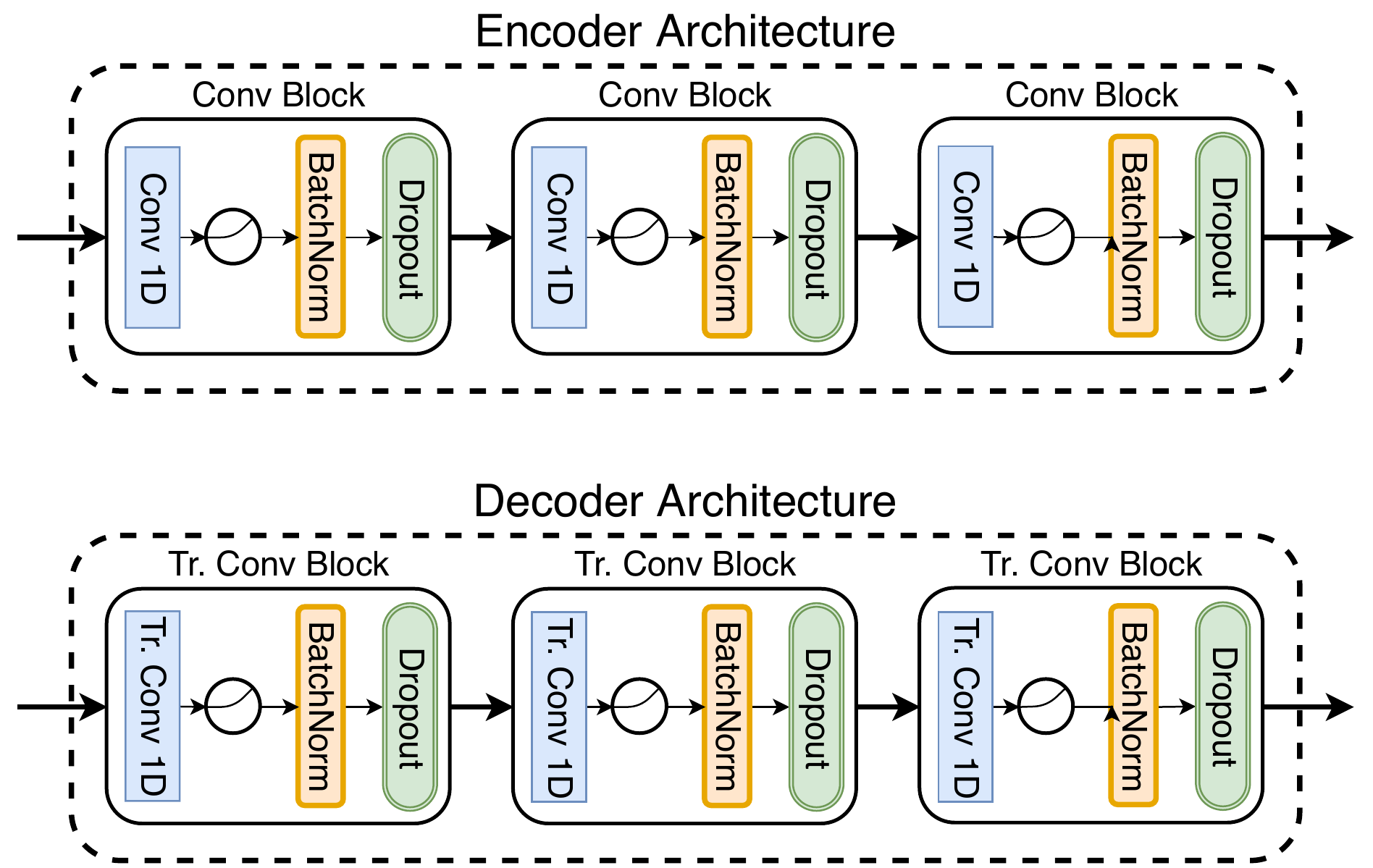}
  \caption{Block diagram of encoders and decoders used for unsupervised domain translation. The convolution and transposed convolution layers are followed by a softplus non-linearity.}~\label{fig:enc_dec_block_diagram}
\end{figure}

\subsubsection{Encoder/Decoder Architecture}
\label{sssec:enc_dec_architecture}
Figure~\ref{fig:enc_dec_block_diagram} shows a block diagram of the encoder and decoder architectures used in this paper. For the encoders, the input is a spectorgram with $1024$ frequency coefficients. The encoder is formed by a cascade of three conv-layer blocks. Each conv-layer block consists of a 1D convolutional layer with a kernel width of $5$ and a soft-plus non-linearity, followed by a batch normalization layer~\cite{ioffe2015batch} and a dropout layer with $p=0.3$~\cite{srivastava2014dropout}. Each conv-layer block transforms its input into a $1024$-dimensional space ($1024$ input and output channels). The final conv-layer block gives the latent representation. Similarly, the decoder architecture consists of a series of three transposed conv-layer blocks. The transposed conv-layer block is identical to the conv layer block except for the 1D convolutional layer which is now replaced by a 1D transposed convolutional layer.

\subsection{Cost Function}

\label{ssec:cost_function}
\subsubsection{Reconstruction Loss}
\label{sssec:reconstruction_loss}
The network architecture basically consists of two auto-encoders, for the source and target domain respectively. Thus, the cost-function should include a term to minimize the discrepancy between the reconstructions and the inputs. We use the mean squared error (MSE) between $\mathbf{M}$ and $\widehat{\mathbf{M}}$ for the source domain VAE and MSE between $\mathbf{C}$ and $\widehat{\mathbf{C}}$ for the target domain VAE. 

\subsubsection{Consistency Loss}
\label{sssec:consistency_loss}
To ensure that the two auto-encoders are not trained independently and share a common latent representation, a necessary condition is to include the following cycle-consistency terms into the cost-function~\cite{liu2017unsupervised, zhu2017unpaired, kaneko2018cyclegan}. The consistency terms ensures that for a point $h$ in the latent space, a decoding followed by an encoding operation maps to vicinity of the same point in the latent space. We use two forms of cycle-consistency terms in the loss function. For the straight-consistency terms, we ideally want, 

\begin{align*}
  \mathbf{\mathcal{E}_{s}}( \mathbf{M}) &= \mathbf{\mathcal{E}_{s}}( \mathbf{\mathcal{D}_{s}}( \mathbf{\mathcal{E}_{s}}( \mathbf{M}))) \\ \mathbf{\mathcal{E}_{t}}( \mathbf{C}) &= \mathbf{\mathcal{E}_{t}}( \mathbf{\mathcal{D}_{t}}( \mathbf{\mathcal{E}_{t}}( \mathbf{C})))
\end{align*}
Similarly, for the cross-consistency terms we require,
\begin{align*}
  \mathbf{\mathcal{E}_{s}}( \mathbf{M}) &= \mathbf{\mathcal{E}_{t}}( \mathbf{\mathcal{D}_{t}}( \mathbf{\mathcal{E}_{s}}( \mathbf{M}))) \\ \mathbf{\mathcal{E}_{t}}( \mathbf{C}) &= \mathbf{\mathcal{E}_{s}}( \mathbf{\mathcal{D}_{s}}( \mathbf{\mathcal{E}_{t}}( \mathbf{C})))
\end{align*}
We measure the discrepancy between the left and right sides of the above equations using their MSEs and incorporate them into the cost-function for minimization. \\

Finally, as a consequence of using a VAE, we also have a $l_{2}$ sparsity term on the encoder outputs to push the latent representation towards a zero-mean distribution. The overall cost-function is a combination of the reconstruction, cycle-consistency and $l_{2}$ terms.

\begin{figure*}[h]
\centering
  \includegraphics[clip, trim = 0.4cm 2.1cm 0.3cm 1.3cm, width=0.7\linewidth]{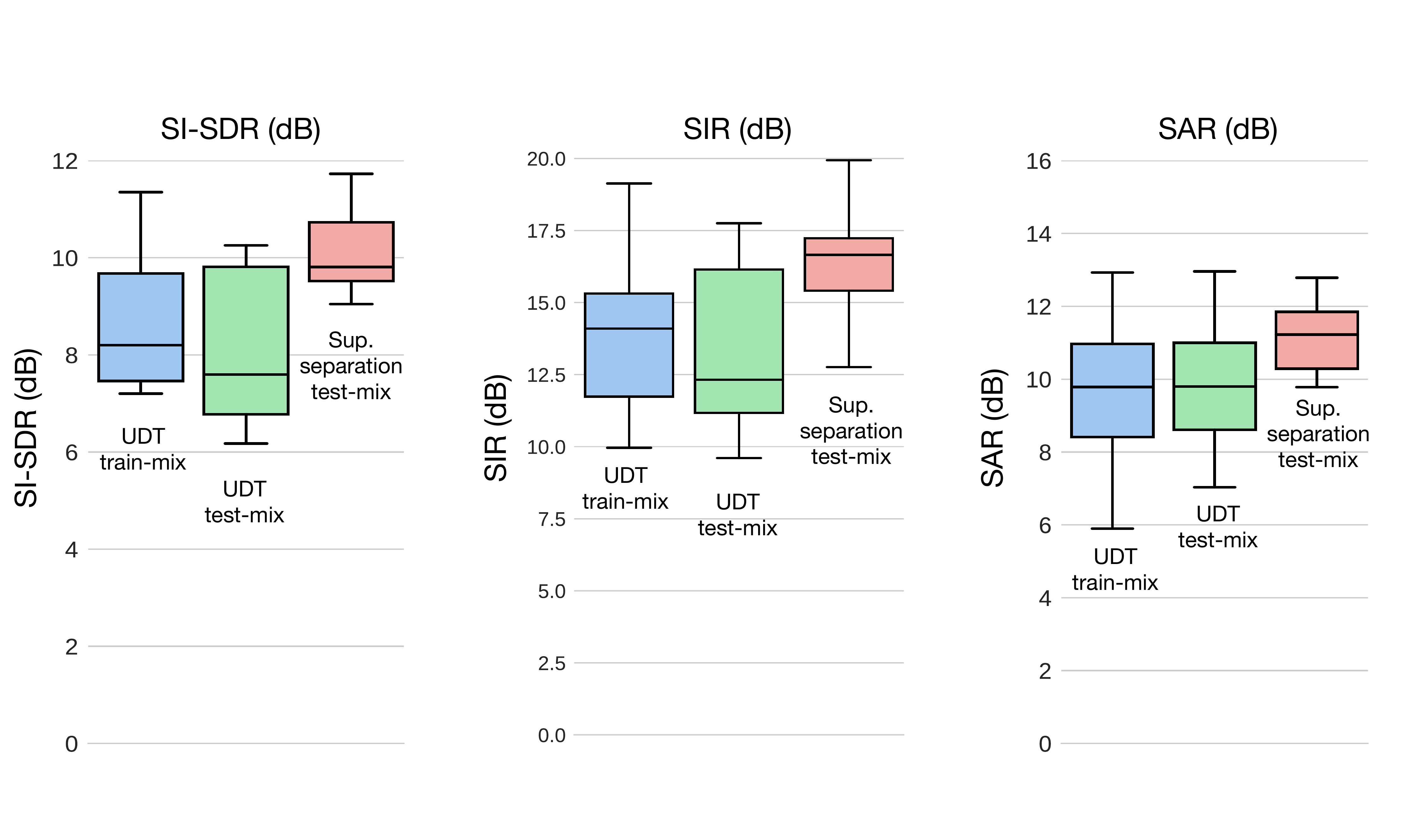}
  \caption{Box plots of SI-SDR, SIR and SAR for UDT on the training set mixtures (left), UDT on the test set mixtures (middle) and supervised separation on the test set mixtures (right). The solid line in the center gives the median value. The extremities of the box show the inter-quartile range ($25^{th}$ and $75^{th}$ percentile points). We see that the separation performance of UDT without paired training examples is not considerably worse than supervised source separation. }~\label{fig:udt_sisdr}
\end{figure*}

\section{Experiments}
\label{sec:experimets}
A key feature of using UDT for source separation is that it allows us to identify and extract a specific source of interest, without any information about the interfering sources in the mixture. We setup a source separation task along these lines to evaluate the separation performance of UDT and compare it with its supervised counterpart. The details and results of the experiments are presented next.

\subsection{Experimental Setup}
\label{ssec:experimental_setup}
In our separation experiment, we train our networks to extract the utterance of the female speaker from a mixture of one male and one female speaker. For the training data, we generate mixture and female-speech examples having a duration of $2$~secs each, using the TIMIT database~\cite{timit}. We use the recordings from ``dr1'', ``dr2'' and ``dr3'' directories for our experiments. Training the UDT network for source separation requires a set of mixture examples and an unpaired set of clean female speech examples. To generate the set of mixtures, we randomly select a pair of male and female speakers from dr1 and randomly pick an utterance for each speaker from the available $10$ recordings. We then randomly select $2$-sec snippets from each selected recording and mix them at $0$~dB to generate the mixture. Thus the mixture examples entirely come from the dr1 directory. For female speech examples, we use the female speakers from dr2. As before, we randomly select female speaker recordings and generate $2$-sec snippets. For the test set, we generate similar (mixture, female-speech) pairs from dr3 and evaluate the networks on their separation performance on this test set. The magnitude spectrograms of these recordings are computed using an STFT operation with a window size of $2048$ samples and a hop of $16$ samples. We raise the $1024$-dimensional magnitude spectrograms by a power of $0.7$ to emphasize the contribution of low-energy time-frequency bins. 

To assess the upper bound of separation performance, we compare the separation performance of UDT with an equivalent supervised source separation model. We use a denoising VAE having the same encoder, decoder architecture as described in Figure~\ref{fig:enc_dec_block_diagram}. Recent experiments have shown that VAE based networks are better than auto-encoders for supervised single-channel source separation~\cite{pandey2018monoaural}. The mixtures from dr1 and their paired female speech examples are also used as the training set for a supervised denoising auto-encoder. The network is evaluated and compared on the test set drawn from dr3. 

The UDT and supervised networks are both trained to estimate the spectrogram of the female speech utterance in the mixture. To transform these estimated spectrograms into their respective waveforms, we undo the power operation, multiply with the mixture phase and invert with the inverse-STFT operation. We compare the separation performance using median scale-invariant source to distortion ratio (SI-SDR) values~\cite{le2019sdr} and BSS\_Eval metrics source-to-interference ratio~(SIR) and sources-to-artifact ratio~(SAR)~\cite{vincent2006performance}. The source-to-distortion ratio~(SDR) values obtained were similar ($0.1-0.2$ dB higher) to the SI-SDR metric and showed similar characteristics in terms of variance and median values. 

\subsection{Results and Discussion}
\label{ssec:results_and_discussion}
Figure~\ref{fig:udt_sisdr} gives the separation performance of UDT and supervised network. The results are shown in the form of a box plot of SI-SDR values.  The middle and right box plots compare UDT and supervised separation on the test set drawn from dr3. Since UDT did not access the clean versions of the mixture examples used for training, we also report the separation performance on the mixture examples used for training. The box plot on the left gives the SI-SDR values on the training set drawn from dr1. We show these plots on a set of $30$ mixtures for each case.

On the test-set, we see that UDT gives a median separation performance of $7.5$~dB, an approximate drop of $2$ dB compared to supervised source separation. The variance of the metrics shown is also higher for UDT. However, these results are comparable and significant considering the fact that UDT learns to separate mixtures without any information about the corresponding clean versions of the sources. The separation performance evaluated on the training mixtures seems to provide an improvement over the test set. Also, we can learn to separate mixtures without any information regarding the other sources in the mixture, along the lines of supervised source separation. We also note that we did not explicitly specify the style and content of our domain transfer problem. This information was implicitly embedded in the training datasets by providing several unpaired examples of mixtures and clean female speech. Similar to partial cofactorization versions of NMF~\cite{kim2011nonnegative}, UDT understands and exploits the common structure across the domain of mixtures and the domain of clean sources to automatically identify the style and content. Thus, UDT can potentially spawn algorithms for other audio style transfer applications.


\section{Conclusion}
\label{sec:conclusion}
Supervised source separation using neural networks requires paired training examples where, we input a mixture into the network and adapt the parameters to produce an output the resembles the corresponding clean source. To ease these requirements, we investigated the use of unsupervised domain (style) transfer for source separation. Learning a shared latent representation for the domain of mixtures and the domain of clean examples allows us to perform separation by mapping between these domains without paired training examples. The separation experiments show that we can extract sources   without a substantial degradation in separation performance compared to supervised source separation. The network also understands the application specific nature of style and content in an audio context from the training examples provided, possibly providing a general framework for other audio style transfer tasks.

\bibliographystyle{IEEEbib}
\bibliography{strings,refs}

\end{document}